\documentclass[12pt]{article} 
\def\be{\begin{equation}} \def\ee{\end{equation}}
\def\bi{\begin{itemize}} \def\ei{\end{itemize}}
\def\bea{\begin{eqnarray}} \def\eea{\end{eqnarray}} \def\ba{\begin{array}}
\def\ea{\end{array}} \def\ben{\begin{enumerate}} \def\een{\end{enumerate}}

\newcommand{\eqn}[1]{(\ref{#1})}

\newcommand{\prl}[3]{Phys. Rev. Lett. {\bf#1} ({#2}) {#3}}

\newcommand{\hepth}[1]{{\tt arXiv:{#1}[hep-th]}}

\def\l{\lambda}

\def\br{\nonumber\\}
 
\def\g{\gamma}

\def\tp{{\tilde{p}}}

\begin{document}
{}~
\hfill \vbox{
\hbox{\today}}\break

\vskip 3.5cm
\centerline{\Large \bf
 Entanglement asymmetry
}
\centerline{\Large \bf
for boosted black branes and the bound
}

\vspace*{1cm}

\centerline{\sc  
Rohit Mishra and Harvendra Singh
}

\vspace*{.5cm}
\centerline{ \it  Theory Division, Saha Institute of Nuclear Physics} 
\centerline{ \it  1/AF Bidhannagar, Kolkata 700064, India}
\vspace*{.25cm}

\vspace*{1.5cm}

\centerline{\bf Abstract} \bigskip
We study the effects of asymmetry 
in the entanglement thermodynamics of CFT subsystems.
It is found that  `boosted'  D$p$-brane backgrounds 
give rise to the first law of the entanglement thermodynamics
where the CFT pressure asymmetry plays a decisive role in the entanglement. 
Two different strip like subsystems, one parallel to the boost and 
the other perpendicular, are studied
 in the perturbative regime  $T_{thermal}\ll T_E$. We mainly seek to
quantify this entanglement asymmetry as a ratio of 
the first order entanglement entropies of the excitations.  
We discuss  the AdS-wave backgrounds at zero temperature having maximum 
asymmetry from where  a bound on entanglement asymmetry is obtained.
The entanglement asymmetry reduces as we switch on finite temperature
in the CFT while it is maximum at zero temperature.

\vfill 
\eject

\baselineskip=16.2pt


\section{Introduction}

The AdS/CFT correspondence \cite{malda} has been a quite
successful idea in string holography. It relates conformal field theries
living on the boundary of  anti de Sitter (AdS)
spacetime with the supergravity theory in the bulk. 
The holographic idea of entanglement entropy
has also been a focus of recent study 
in string theory \cite{RT}.
It has led to some understanding of entanglement entropy in strongly 
coupled quantum mechanical systems, 
particularly  theories which exhibit scaling property near 
the critical points \cite{ogawa}. 
One significant observation has been 
that the small excitations of the subsystem in
the boundary CFT follow  entanglement thermodynamic  laws somewhat
similar to the black hole thermodynamics \cite{JT,alisha},
 see also related works \cite{Pang:2013lpa}, 
\cite{Wong:2013gua}, \cite{Park:2015afa}.
These calculations have become possible now because
 entanglement entropy can be studied by using  gauge/gravity holography 
\cite{RT}, that is
 by evaluating the geometrical area of some spatial extremal sufaces embedded 
inside  asymptotically AdS  geometry. It has been proposed recently 
in \cite{JT} that the entanglement entropy ($S_E$) and 
the energy of excitations (${\cal E}$) in a pure AdS background give rise to
a thermodynamic relation
$$\bigtriangleup { \cal E}
= 
T_E \bigtriangleup S_E  +
{\cal V} \bigtriangleup {\cal P}  +
\mu_E \bigtriangleup N   
$$
This isn described as the first law of entanglement 
thermodynamics \cite{JT,alisha}, while the charge and chemical potential
contributions can  arise  in the boosted black-brane cases \cite{ms2015}.
These charge excitations could be the Kaluza-Klein (momentum) modes along  
the compact circle or  the winding modes of  string wrapped along 
a T-dual circle.

In this paper we particularly  study the effects of boost (excitations)
in asymptotically AdS spacetimes  on the nature of entanglement first law. 
We  find that the boosted  black branes
give rise to an asymmetry in the entanglement first law. 
We study two types of strip  subsystems
one parallel to the boost and the other perpendicular to the boost direction.
There is difference in the `entanglement pressure' in two cases such that
$\bigtriangleup {\cal P}_\perp \le \bigtriangleup {\cal P}_\parallel$.
We find that primarily the entanglement
pressure is responsible for the differences in the entanglement entropies,
$\bigtriangleup S_\perp \ge \bigtriangleup S_\parallel$, in the two cases.
The entanglement asymmetry may be quantified as a dimensionless ratio
\bea
{\cal A}&&\equiv  
{\bigtriangleup S_\perp - \bigtriangleup S_\parallel
\over 
\bigtriangleup S_\perp+\bigtriangleup S_\parallel}\br &&
={\beta^2\gamma^2\over ( 2 + {d+3\over d-1}\beta^2\gamma^2)}
\le {d-1\over d+3}\nonumber
\eea
We find that the asymmetry depends only on boost and it is bounded from above.
The bound is saturated
only for the AdS-wave background, which is the case involving infinite boosts.
To obtain these results we  resort to a  perturbative 
calculation of the entanglement entropy up to first order, where
 the ratio ${l \over z_0}$, of 
the strip width ($l$) to the horizon size 
($z_0$), is kept very small. This 
hierarchy of length scales can also be thought of in  terms of the
temperatures as  
${T_{thermal}\over T_E} \ll {a_1\over 2b_0\gamma} \ .$

The paper is   organized as follows. 
In the second section we first reproduce our earlier results for 
the perpendicular strip case at first order in the perturbative expansion.
Then we  take up the case when the strip subsytem is parallel to the boost
direction. The entanglement laws are obtained at the first order. 
In section-3 we define entanglement  asymmetry ratio. 
It is found that it  depends only on the boost velocity, 
which is a measurable effect. 
We also obtain entanglement asymmetry in the AdS-wave case also, in which case
asymmetry saturates the upper bound. We extend 
these results to nonconformal D-brane case also in section-4. 
Finally we conclude in  section-5.

\section{Entanglement from boosted black-branes}

The boosted $AdS_{d+1}$ backgrounds we are interested are
given by
\bea\label{bst1}
&&ds^2={L^2\over z^2}\left( -{f dt^2\over K}+K (dy-\omega)^2
+dx_1^2+\cdots+dx_{d-2}^2+{dz^2 \over f}\right) 
\eea
with functions
\be
 f=1-{z^d\over z_0^d}, ~~~~~K=1+\beta^2\gamma^2{z^d\over z_0^d} 
\ee
 $z=z_0$ is the horizon and  $0\le \beta\le 1$ is  boost parameter, while
 $\gamma={1\over\sqrt{1-\beta^2}}$. The boost is taken along $y$ direction.
The one-form 
\be
\omega={\beta^{-1}}(1-{1\over K}) dt
\ee
and  $L$ is the radius of curvature of  AdS spacetime, which is taken 
very large in string length units. 
\footnote{
For example, in the $AdS_5\times S^5$ 
near-horizon geometry of $n$ coincident D3-branes, 
we shall have $L^4\equiv 2\pi g_{YM}^2 n$ and
the 't Hooft coupling constant $g_{YM}^2 n \gg 1$.}   
\subsection{A thin (perpendicular) strip }

We first study the entanglement entropy law for a subsystem on the 
boundary of the $AdS_{d+1}$ backgrounds  \eqn{bst1}
where strip is perpendicular to the boost direction:
the strip width is $-l/2 \le x^1 \le l/2$, while 
the boost is along $y$ direction.
Thus the steps in this section are same as in our presious work \cite{ms2015}.
We  embed  the $(d-1)$-dimensional strip-like  (constant $t$ surface)
inside  the bulk  geometry. 
The two boundaries of the
extremal  surface   coincide with the two ends 
of the interval $\bigtriangleup x^1$.   
The  size of the
rest of the coordinates, 
$0\le y\le l_y$, 
$0\le x^i\le l_i$, 
is taken very large, such that $l_y, l_i\gg l$. 
As per the Ryu-Takayanagi prescription \cite{RT}
the entanglement entropy of the strip subsystem  is given 
in terms of the geometrical area of the extremal surface
(constant time) 
\bea\label{schkl1saa}
 S_\perp \equiv {[A]_{Strip}\over 4G_{d+1}} =
 { V_{d-2}  L^{d-1} \over 
2G_{d+1}}
\int^{z_\ast}_{\epsilon}{dz\over z^{d-1}} \sqrt{K} 
\sqrt{{1\over {f}}  +({\partial_z x^1})^2}
\eea  
where  
$G_{d+1}$ is $(d+1)$-dimensional Newton's constant (of bulk gravity)
and $V_{d-2} \equiv  l_y l_2 l_3 \cdots l_{d-2}$ is 
the net spatial volume of the strip  on the boundary. 
We will be mainly working for $d>2$ here.
In our notation $z=\epsilon\sim 0$ is the cut-off scale
and $z=z_\ast$ is the turning point
of extremal surface. 
In the above area functional
$K(z)$, and $f(z)$
are known functions, so we only need to extremize for $x^1(z)$. 
After extremization the entanglement entropy
for perpendicular strip subsytem can be written as 
\be\label{kl1kv}
S_\perp=
{  V_{d-2} L^{d-1} \over 2 G_{d+1}}
\int^{z_\ast}_{\epsilon}{dz \over  z^{d-1}}{K 
\over\sqrt{f}\sqrt{K- K_\ast({z\over z_\ast})^{2d-2}}} 
\ee  
where  $K_\ast\equiv K(z)|_{z=z_\ast}$.
The  boundary value $x^1(0)=l/2$ has the integral
relation 
\be\label{klop}
{l \over 2}=  \int_0^{z_\ast} dz
({z\over z_\ast})^{d-1}{1 \over \sqrt{f} 
 \sqrt{{K\over K_\ast}- ({z\over z_\ast})^{ 2d-2}}} 
\ee
which relates $l$ with  the turning point $z_\ast$.   
The turning-point takes the mid-point value $x^1(z_\ast)=0$ on the boundary. 

When  strip subsystem is a small 
the turning point will lie in the proximity of
 asymptotic boundary region only ($z_\ast \ll z_0$).  
We can evaluate the entanglement entropy 
\eqn{kl1kv} by expanding it around the AdS  (i.e. treating pure AdS
as a ground state). We  take boost to be finite such that 
\be\label{dfg0a}
 {z_\ast^d \over z_0^d} \ll 1,~~~~ 
 {(\beta\gamma)^2 z_\ast^d\over z_0^d} 
\ll 1 
\ee
is always maintained. In this limit we can estimate the entropy perturbatively.
Under these approximations, 
entanglement entropy contribution (above pure AdS) at first order
is given by \cite{ms2015} \footnote{
The coefficients  (given in \cite{ms2015} also) are defined as;
$
 b_0=\int_0^{1} d\xi \xi^{d-1}{1 \over   \sqrt{1-\xi^{2d-2}}}
={1\over 2(d-1)}B({d\over2d-2},{1\over2})\equiv (2-d)a_0 ,~~~
 b_1=\int_0^{1} d\xi \xi^{2d-1}{1 \over   \sqrt{1-\xi^{2d-2}}}
={1\over 2(d-1)}B({d\over d-1},{1\over2})\equiv{2\over d+1} a_1
$, 
where $B(m,n)={\Gamma(m)\Gamma(n)\over\Gamma(m+n)}$ are the Beta-functions.
}
\bea\label{hj4}
&& \bigtriangleup 
S_\perp = 
S_\perp  -
S_{AdS}  
= {L^{d-1} V_{d-2}\over 16 G_{d+1}} {a_1 l^2\over  b_0^2}
 \left(
{d-1\over d+1}+\beta^2\g^2 
 \right) {1\over  z_0^d} \ .
\eea
The CFT energy and  pressure 
 can be obtained by expanding the bulk geometry \eqn{bst1} in
 Fefferman-Graham  coordinates valid near the  boundary \cite{fg}. 
The energy of the excitations is 
\bea\label{hj5}
&& \bigtriangleup {\cal E}
=  {  L^{d-1} V_{d-2}l \over 16\pi G_{d+1}}  
({d-1\over d} +\beta^2\gamma^2) {d\over
z_0^d}
\eea
The volume  is $V_{d-2}\equiv l_y l_2 \cdots l_{d-2}$.
The pressure along $y$ direction is
\bea
&& 
\bigtriangleup {\cal P}_\parallel=
\bigtriangleup {\cal P}_y
= { L^{d-1}  d \over 16 \pi G_{d+1}}  
({1\over d}+\beta^2\gamma^2){1 \over z_0^d}
\eea
while the 
pressure  along all other $x_i$'s (perpendicular to the boost  direction)  is
identical and is given by
\bea\label{hj5a}
&& 
\bigtriangleup {\cal P}_\perp
= {L^{d-1} \over 16\pi  G_{d+1} }  {1 \over z_0^d}= \bigtriangleup {\cal P}_1
=\bigtriangleup {\cal P}_2=
\cdots 
\eea
This pressure asymmetry is solely due to the boost. For example
the pressure is more along the
$y$ (boost) direction as compared to  $x^i$'s coordinates.  
Using \eqn{hj5} and \eqn{hj5a} we can  express eq.\eqn{hj4} as
\bea \label{alis1}
&& \bigtriangleup S_\perp  
= {1\over T_E} (
\bigtriangleup {\cal E}-
{d-1\over d+1} ~{\cal V}_\perp
\bigtriangleup {\cal P}_\perp ) 
\eea
where $ {\cal V}_\perp
\equiv l [l_yl_2\cdots l_{d-2}] $ is the net volume of the strip subsystem. 
The entanglement temperature is given by
\be\label{temp1}
T_E^\perp= 
{ (B({d\over 2d-2},{1\over2}))^2 \over 2(d-1) 
B({1\over d-1},{1\over2})}{d\over \pi l}.
\ee
The temperature
 is inversely proportional to the width of strip. 
The equation \eqn{alis1} simply
 describes the  first law of entanglement thermodynamics \cite{alisha,JT}. 
Subtle changes will occur in  this  expression 
when strip is taken along the boost.

\subsection{Strip  along the boost} 

We now study the entanglement entropy of a strip 
subsystem such that its width is
parallel to the boost (flow) direction. That is,  we take
the  boundaries of the
extremal surface to coincide with the two ends 
of  $\bigtriangleup y$ interval: $-l/2 \le y \le l/2$.   
The regulated  size of 
rest of the coordinates will be taken much larger
$0\le x^i\le l_i$, such that $l_i\gg l$ $(i=1,2,\cdots, d-2)$.\footnote{
We wish to embed the $\bigtriangleup y$ interval, but since
 the boost is also along $y$, both `time' $t(z)$ and  $y(z)$ 
would have to be embedded in the bulk in a covariant manner 
\cite{Hubeny:2007xt}. So one 
has to be a bit cautious while working with stationary metric cases 
\cite{Nozaki:2013vta} \cite{Blanco:2013joa}.  
However, it can be explicitly shown that, in the perturbative expansion
(for small strips) 
to know the  entropy  only upto first order (next to the pure AdS), 
just taking a constant $t$  slice would suffice. The deviations in  extremal
surface geometry away from the constant time slice will contribute only 
to the second order terms in the expansion.
Our aim in this work is to know   only the first order
terms in the expansions of $z_\ast$ and strip area.} 
 Taking the constant time slice the 
 entanglement entropy  of the parallel strip becomes
\bea\label{schkl1saapar}
 S_\parallel  
&=&
 { V_{d-2}  L^{d-1} \over 
2G_{d+1}}
\int^{z_\ast}_{\epsilon}{dz\over z^{d-1}} 
\sqrt{{1\over {f}}  +K({\partial_z y})^2}
\eea  
where  now 
 $V_{d-2} \equiv l_1 l_2  \cdots l_{d-2}$ is 
the  spatial volume. 
The identification of the extremal 
strip boundary, $y(0)=l/2$, leads to the integral
relation 
\be\label{kloppar}
{l \over 2}=  \int_0^{z_\ast} dz
({z\over z_\ast})^{d-1}{1 \over \sqrt{f K} 
 \sqrt{{K\over K_\ast}- ({z\over z_\ast})^{ 2d-2}}} 
\ee
which relates $l$ with  the turning point $z_\ast$ of the strip.   
The turning-point takes the mid-value $y(z_\ast)=0$. 
The final expression of the entanglement entropy
for the parallel strip subsystem  now becomes
\be\label{kl1kvpar}
S_{\parallel}=
{  V_{d-2} L^{d-1} \over 2 G_{d+1}}
\int^{z_\ast}_{\epsilon}{dz \over  z^{d-1}}{\sqrt{K} 
\over\sqrt{f}\sqrt{K- K_\ast({z\over z_\ast})^{2d-2}}} 
\ee  
Since the parallel system has not been covered in \cite{ms2015}
let us provide some essential details perturbative calculation
here. In small strip cases,
the   equation \eqn{kloppar} can be expanded perturbatively as
\bea \label{dfg1plpar}
{l}=
2z_\ast \left( b_0 + 
{z_\ast^{d}\over 2 z_0^d}( (1+{2\beta^2\gamma^2\over d-1}) b_1 
- {\beta^2\gamma^2\over d-1} b_0 )\right)  
+\cdots
\eea
where dots indicate terms of higher powers in 
$({z_\ast\over  z_0})^{d}$, and various
 coefficients are defined earlier. From here
keeping only up to  first order 
the equation implies
\bea\label{dfg5par}
 z_{\ast}&=& {\bar z_\ast \over  
1 + {\bar z_\ast^{d}\over 2 z_0^d}( 
(1+{2\beta^2\gamma^2\over d-1}) {b_1\over b_0} 
- {\beta^2\gamma^2\over d-1} )} 
\eea
where  
$\bar z_\ast\equiv {l  \over 2 b_0}$  being the turning point 
of  pure AdS having the same strip width as $l$. 
Having obtained the turning point expansion,
a similar expansion around  pure AdS  
can be made for the  area functional also. Suppressing the details,
after regularizing the area integral  \eqn{kl1kvpar}, 
the net change in the area of parallel strip (above  pure AdS 
value)  comes out to be
\bea
 \bigtriangleup A_\parallel&=&
 {a_0\bar z_\ast^2 \over  z_0^d}(  {a_1\over a_0} 
- (1-\beta^2\g^2){b_1 \over a_0} )  
\eea
and corresponding change in the entropy for parallel strip  becomes
\bea\label{hj4par}
&& \bigtriangleup S_\parallel  
= 
{L^{d-1} V_{d-2}\over 16 G_{d+1}} {a_1 l^2\over  b_0^2}
 \left(
{d-1\over d+1}+{2\over d+1} \beta^2\g^2 
 \right) {1\over  z_0^d} .
\eea
The equation \eqn{hj4par}
is   complete  expression up to the first order.
The entanglement first law for a strip along
the flow becomes
\bea\label{hj4par1}
&& \bigtriangleup S_\parallel  
= {1\over T_E^\parallel}(
\bigtriangleup E_\parallel -{d-1\over d+1} 
{\cal V}_\parallel 
\bigtriangleup P_\parallel) 
\eea
where ${\cal V}_\parallel=l V_{d-2}=l [ l_1 l_2\cdots l_{d-2}]$,
and $\bigtriangleup {\cal P}_\parallel=\bigtriangleup {\cal P}_y$
is defined earlier. The temperature is
\be\label{temp1par}
T_E^\parallel= { b_0^2\over a_1}{d\over \pi l}=T_E^\perp.
\ee
The two temperatures remain the same but the entanglement entropies differ
significantly.

\section{Entanglement asymmetry  and the bound}

Following from  previous section,
with out any loss of generality we
 can always take the volume of the strip subsystems to be equal 
\be
{\cal V}_\parallel= {\cal V}_\perp=l. V_{d-2}.
\ee
This only means that regulated size of the boxes is kept 
the same in both the cases,
along with the strip width  
$l$. It  implies that
 \be
T_E^\parallel=T_E^\perp, ~
\bigtriangleup {\cal E}_\parallel=\bigtriangleup {\cal E}_\perp.
\ee
Comparing the two types of entropy results,  the difference is
given by
\bea\label{opl8}
\bigtriangleup S_\perp-\bigtriangleup S_\parallel &=&
{L^{d-1} V_{d-2}\over 16 G_{d+1}} {a_1 l^2\over  b_0^2}
 \left({d-1\over d+1} \beta^2\g^2 
 \right) {1\over  z_0^d}\br
&& ={d-1\over d+1} ~{\cal V}(
\bigtriangleup {\cal P}_\parallel - 
\bigtriangleup {\cal P}_\perp ) 
 \ .  
\eea
The right hand side is a positive definite expression. Hence
we can deduce that entanglement entropy 
is more for a perpendicular 
strip subsytem as compared to  the parallel set-up, 
eventhough the energy of  excitations and 
 entanglement temperatures remain the same for both. The key to this 
entropy enhancement effect, 
\be
\bigtriangleup S_\perp\ge \bigtriangleup S_\parallel \ee
can directly  be alluded to unequal entanglement pressure;
\be
\bigtriangleup {\cal P}_\perp\le \bigtriangleup {\cal P}_\parallel .\ee
Thus more energy is consumed by the excitations in the parallel strip
(with an increased pressure)
 as compared to the perpendicular strip
(having a low pressure along the strip). This suggests that in
the boundary  CFT `pressure' plays a vital role in determining the 
entanglement entropy of the subsytems.  
The equation \eqn{opl8} also implies that, up to  first order,
the net difference of the entanglement entropies is
\bea\label{opl8a}
 S_\perp- S_\parallel &&=
{L^{d-1} V_{d-2}\over 16 G_{d+1}} {a_1 l^2\over  b_0^2}
 \left({d-1\over d+1} \beta^2\g^2 
 \right) {1\over  z_0^d}
\eea
Thus the entropy asymmetry coexists with
 pressure asymmetry  in the CFT, whereas
$\bigtriangleup S_\perp|_{\beta=0}=
\bigtriangleup S_\parallel|_{\beta=0}.
$

We can now define the entanglement asymmetry as a ratio
\be\label{asym56}
{\cal A}\equiv  
{\bigtriangleup S_\perp - \bigtriangleup S_\parallel
\over 
\bigtriangleup S_\perp +\bigtriangleup S_\parallel}
={\beta^2\gamma^2\over ( 2 + {d+3\over d-1}\beta^2\gamma^2)}
\ee
Thus  nonzero boost ($\beta\le 1$) will always induce 
  entanglement asymmetry in the boundary CFT. 
The asymmetry will however vanishes for $\beta=0$.
Note that these results 
have been derived in the perturbative regime described in \eqn{dfg0a} 
only up to  first order. We also learn that
the asymmetry will always be bounded. 
In the above the bound is saturated only
in the large boost limit, which we shall discuss in the next section.

We could however define an entanglement entropy ratio as
\be\label{jk89}
{\cal R}\equiv  
{\bigtriangleup S_\parallel 
\over \bigtriangleup S_\perp}= 
{1 +{2\over d-1} \beta^2\g^2 \over
1+{d+1\over d-1} \beta^2\g^2} 
\ge 
{2\over d+1} 
\ee
a quantity which  depends on the  boost only and is devoid of external 
factors like shape and size. Then
\be
{\cal A}\equiv  {1 -{\cal R} \over 1 +{\cal R} } 
\le 
{d-1\over d+3}. 
\ee
We shall show that
the bound is saturated in the case of AdS-wave in the next section.
The maximum value ${\cal R}$ can take is one for which entanglement asymmetry 
vanishes.

\subsection
{$\beta\to 1, ~z_0\to\infty$ limit (pressureless system)}

In  present examples
the pressure in the CFT$_d$ can be controlled by regulating the boost. 
We now show that there exists a simultaneous double limit in which 
the pressure asymmetry of the CFT excitations becomes optimal.
We take a double limit $\beta\to 1, ~z_0\to\infty$,  keeping the
ratio
\be\label{dl2}
{\beta^2\gamma^2 \over z_0^d}={1\over z_I^d}={\rm Fixed}\ .
\ee
These double limits has previously been explored in \cite{Singh:2010zs}
 in connection with 
Lifshitz type backgrounds from black D$p$ branes (lightcone coordinates). 
Under these limits
the bulk geometry \eqn{bst1} reduces to the following  AdS-wave background 
\bea\label{bstpar}
&&ds^2={L^2\over z^2}\left( -K^{-1}{ dt^2}+K (dy-(1-K^{-1}) dt)^2
+dx_1^2+\cdots+dx_{d-2}^2+{dz^2  }\right) 
\eea
with the new function $K=1+{z^d\over z_I^d}$, where
 $z=z_I$ is an  scale which determines momentum of the wave travelling
in the $y$ direction. (The entanglement of strip systems 
 for AdS-waves has previously been explored by \cite{narayan} also.) 
For this background the energy of the excitations in the CFT becomes
\bea\label{hj5pas}
&& \bigtriangleup {\cal E}=
  {  L^{d-1} V_{d-2}l \over 16\pi G_{d+1}}   {d\over
z_I^d}
\eea
The pressure along the wave ($y$) direction is
\bea
&& 
\bigtriangleup {\cal P}_\parallel=
\bigtriangleup {\cal P}_y=
 { L^{d-1}  d \over 16 \pi G_{d+1}}  {1 \over z_I^d}
\eea
while the 
pressure  along all $x_i$'s (perpendicular to the wave direction)  
identically vanishes
\be
\bigtriangleup {\cal P}_\perp=0
\ee
 in the  boundary CFT$_d$, which is a conformal theory with 
traceless energy-momentum tensor. 

The double limits can  be directly employed on the  entropy
results  obtained in the previous section, provided
we maintain ${z_\ast^d\over z_I^d}\ll 1$. 
Employing the limits on the entropy
expressions in eqs. \eqn{hj4} and \eqn{hj4par}, it gives us
 \bea\label{hj4par3}
&& \bigtriangleup S_\perp
= {L^{d-1} V_{d-2}\over 16 G_{d+1}} {a_1 l^2\over  b_0^2}
  {1\over  z_I^d} ={1\over T_E}(
\bigtriangleup {\cal E} )
\eea
while
\bea\label{hj4par4}
&& \bigtriangleup S_\parallel 
= 
{L^{d-1} V_{d-2}\over 16 G_{d+1}} {a_1 l^2\over  b_0^2}
 \left( {2\over d+1} 
 \right) {1\over  z_I^d}= {1\over T_E}(
\bigtriangleup {\cal E}
 -{d-1\over d+1}{\cal V}\bigtriangleup {\cal P}_\parallel)
\eea
The  width $l$ is kept the same in both cases as well as the volumes. Hence
 entanglement temperatures, $
T_E= { b_0^2\over a_1}{d\over \pi l}$, and 
$\bigtriangleup{\cal E}$,  remain the same for both the cases. 
Particularly in the former case there is no  entanglement pressure 
along the strip ($x^1$ direction). As no `entanglement work' seems
to have been done by the excitations  due to vanishing pressure, 
 the entropy remains  maximal in the perpendicular direction.
\footnote{ The `entanglement work' or  aptly the `work function'
could be defined as: 
$\bigtriangleup {\cal W}=l{ V_{d-2}}\bigtriangleup {\cal P}=l \cdot{\cal F}$.
The physical origin of entanglement work may be traced to the 
fact that finite pressure causes some energy for the excitations
to work against it. It is always proportional to the size of the strip.
In other words, there is a force ${\cal F}$ on the strip walls, and
 it will take $\bigtriangleup {\cal W}$ amount of energy in
moving  the walls  upto a separation $l$.
This in turn implies there will be a 
 suppression  of the excitations that can
take part in the entanglement. Note
these entanglement quantities  are  purely of non-thermal in nature,
as  thermal temperature is vanishing for the AdS-wave,
correspondingly  thermal entropy and thermal pressure are also vanishing.} 
 While in the latter case
 there is finite pressure along the strip width, so finite energy is consumed
by the excitations to work against the pressure as 
they take part in the entanglement. 
Thus the work done against entanglement pressure costs finite energy 
which essentially leads to a reduction in
the net entanglement entropy in direction parallel to propagation of the wave. 
 
From equations \eqn{hj4par3} and \eqn{hj4par4}
 for the AdS-wave case the ratio becomes
\be\label{hj3par5}
{\cal R}_{wave}=
{ \bigtriangleup S_\parallel \over \bigtriangleup S_\perp 
}={2\over d+1} 
.
\ee
This is a remarkable relation and is identical to one in \eqn{jk89}. It remains
true at the linear order in perturbation (over and above the AdS background). 
At the higher orders in ${z_\ast^d\over z_I^d}$ expansion this result might 
change, we hope to report on this in our future work \cite{ms2016}. 
The entanglement asymmetry  becomes 
\be\label{hj4par5a}
{\cal A}_{wave}\equiv  
{\bigtriangleup S_\perp - 
\bigtriangleup S_\parallel
\over 
\bigtriangleup S_\perp+\bigtriangleup S_\parallel
}={d-1\over d+3} .
\ee
The asymmetry has optimal value and is universal in nature.
The relations \eqn{hj3par5} and \eqn{hj4par5a} are
 applicable only when $d>2$, because
for $d=2$ (i.e. $AdS_3$-wave) the analogue of $\bigtriangleup S_\perp$ 
does not exist, 
but the form of entanglement  first law as in \eqn{hj4par4} 
for  parallel strip does hold good.

\section{Non-conformal boosted black D-branes}
The conformal cases of $AdS$ seometries which are near horizon geometries of 
D3 and M2/M5 branes are covered in the previous section.
In this section we wish to extend entanglement asymmetry analysis
 to the  nonconformal  D$p$ brane backgrounds \cite{itzhaki}.  
We are interested in the boosted  D$p$-brane geometry so that suitable
asymmetry is generated. These nonconformal backgrounds can be written as 
\bea\label{bbst1}
&&ds^2=g_{eff}
\bigg[ -{f \over z^2K}dt^2+{K \over z^2}(dy-\omega)^2
+{dx_2^2+\cdots+dx_{p}^2\over z^2}+{4\over (5-p)^2}{dz^2 \over z^2 f} +
d\Omega^2_{8-p}\bigg]\br
&& e^\phi={(2\pi)^{2-p}\over d_p N}  g_{eff}^{7-p\over 2}
\eea
along with appropriate $F_{(p+2)}$ form Ramond-Ramond flux. The
 strength of the string
 coupling depends on effective YM coupling
$g_{eff}=(\l_p z^{3-p})^{1\over 5-p}$ and
the functions are defined as
\bea
&& f=1-{z^{\tilde p}\over z_0^{\tilde p}}, ~~~~
K=1+\beta^2\gamma^2{z^\tp\over z_0^\tp} \br
&& \omega={\beta^{-1}}(1-{1\over K}) dt
\eea
with  $z=z_0$ being the location of horizon and  $0\le \beta\le 1$ is the boost.
 The boost is taken along the $y$ direction and geometry along brane directions
has  asymmetry. The parameters are defined  as
\be
\l_p\equiv d_p g_{YM}^2 N, ~~\tilde p={14-2p\over 5-p}
\ee
where $d_p$ is a fixed normalisation factor for a given $p$ brane 
(The exact expression will not be needed here but it 
can be found out in \cite{itzhaki}). The parameter
$\l_p$ is essentially the 't Hooft coupling constant and
it controls the curvature of  spacetime which is to be taken
small in string length units ($l_s=1$) and for which $N$ is taken to be
large enough.
The boosted geometry \eqn{bbst1} is conformally  $AdS_{p+2}\times S^{8-p}$,
a near-horizon geometry of $N$ coincident D$p$-branes. Only for $p=3$ case
the geometry becomes conformal and is discussed earlier. 
We are discussing the asymmetry cases for that we need $p=2$ or 
$p=4$, for them at least two asymmetric brane directions are available.

\subsection{Entropy  of thin  strips}

We first consider a thin strip in a
 perpedicular direction to the boost, say $x_2$. 
The Ryu-Takayanagi entropy functional is given by

\bea\label{bschkl1saa}
&& S_\perp =
 {V_{p-1}\Theta_{8-p}Q_p\over 2 G_{N}}
\int^{z_\ast}_{\epsilon}{dz\over z^{\tp-1}} \sqrt{K} 
\sqrt{{4\over (5-p)^2} {1\over {f}}  +({\partial_z x_2})^2}\br
&&= {V_{p-1}\Theta_{8-p}\over 2 G_{N}}
 {2Q_p\over 5-p}
\int^{z_\ast}_{\epsilon}{dz\over z^{\tp-1}} \sqrt{K} 
\sqrt{ {1\over {f}}  +({\partial_z \bar x_2})^2}
\eea  
where $Q_p\equiv {(2\pi)^{2p-4}\sqrt{\l_p^\tp}\over g_{_{YM}}^4}$ while
$\Theta_{8-p}$ is the volume of unit radius $S^{8-p}$ and $ G_{N}$ is
the 10-dimensional Newton's. 
 We shall consider  a small legth interval $-{l\over 2} \le \bar x_2 \le {l\over 2}$, 
but due to the scaling $x_2={2\over5-p}\bar{x}_2$ in eq.\eqn{bschkl1saa} 
the actual 
width of the strip is ${2l\over 5-p}$. One can see
that the integrand in the second line in 
\eqn{bschkl1saa} is
 strikingly same as that for the conformal case discussed earlier,
except that parameter $\tp$ can take  fractional values. (For example, 
for D2-branes $\tp={10\over 3}$,
but for D4-branes $\tp={6}$.) So the 
rest of the calculations is straight forward:
Extremizing the area and making a perturbative expansion keeping the
ratio ${l\over z_0}< 1$, as in 
 previous sections.
Avoiding the unnecessary details we quote the result from eq.\eqn{hj4}.
The entanglement entropy of the excitations above the extremality is
\bea\label{hj4y}
&& \bigtriangleup 
S_\perp 
= {V_{p-1}\Theta_{8-p}\over 16 G_{N}}
 {2Q_p\over 5-p}
{a_1 l^2\over  b_0^2}
 \left({\tp-1\over \tp+1}+\beta^2\g^2 
 \right) {1\over  z_0^d} \
\eea
where new beta functions are given as $b_0\equiv 
{1\over 2(\tp -1)} B({\tp\over 2\tp-2},{1\over2}))$
and $a_1\equiv {1 \over 2(\tp-1)} 
B({1\over \tp-1},{1\over 2})$.
We come to conclusion that the entropy of excitations
in a nonconformal $(p+1)$-dimensional theory 
at the first order can be written as 
\bea \label{alis1n}
&& \bigtriangleup S_\perp  
= {1\over T_E^\perp} (
\bigtriangleup {\cal E}-
{\tp-1\over \tp+1} ~{\cal V}_\perp
\bigtriangleup {\cal P}_\perp ) 
\eea
where $ {\cal V}_\perp={2\over 5-p}l V_{p-1}$
is the net volume of the strip subsystem, while the energy and pressure
 expressions are in appendix. 
The entanglement temperature is defined by
\be\label{temp19j}
T_E^\perp=  
{ (B({\tp\over 2\tp-2},{1\over2}))^2 \over 2(\tp-1) 
B({1\over \tp-1},{1\over 2})}{(7-p)\over  \pi l}.
\ee
The temperature
 is inversely proportional to the width of strip. 
But compared to the law in \eqn{alis1}  
subtle changes have occured in   the pressure term in \eqn{alis1n}.
Namely the coefficient $\tp-1\over\tp+1$ in \eqn{alis1n} is different
from the ratio $ d-1\over d+1$ which appears in \eqn{alis1}. 
(Note $d$  takes only integer values and
is directly correlated with the dimensionality of $AdS_{d+1}$. 
This cannot be said about $\tp$.)  
Let us comment here that for unboosted nonconformal
D-brane case the result 
\eqn{alis1n} was first obtained in \cite{Pang:2013lpa}. So it is interesting
to observe that the form of first law with boost excitations
remains the same as in  unboosted case \cite{Pang:2013lpa}, 
although all physical quantities have themselves changed.

In the next we  consider an strip interval in the direction
parallel to the boost, i.e. along $y$ direction. 
The  entropy functional is given by 
\bea\label{bschkl1saapar}
 S_\parallel  
&=&
{V_{p-1}\Theta_{8-p}\over 2 G_{N}g_{_{YM}}^4}
 {2Q_p\over 5-p}
\int^{z_\ast}_{\epsilon}{dz\over z^{\tp-1}} 
\sqrt{{1\over {f}}  +K({\partial_z \bar y})^2}
\eea  
where  now  $V_{p-1}$ is  regulated  volume of all the $x^i$ coordinates.
We have scaled  $y={2\over 5-p}\bar y$ and
 taken the  width to be $-{l/2}\le \bar y \le {l/2}$.
 As usual extremizing the strip area and expanding up to first order
in the ratio $l/z_0\ll 1$,
we come to conclusion that the entropy of excitations above extremality
for a parallel strip follows the law
\bea \label{alis1lk}
&& \bigtriangleup S_\parallel  
= {1\over T_E^\parallel} (
\bigtriangleup {\cal E}-
{\tp-1\over \tp+1} ~{\cal V}_\parallel
\bigtriangleup {\cal P}_\parallel ) 
\eea
where $ {\cal V}_\parallel={2l\over 5-p} V_{p-1}$
is the net volume of the parallel strip subsystem. 
Since we have kept the same width ${2l\over 5-p}$ in both the situations, 
the entanglement temperature are identical
\be
T_E^\parallel
=T_E^\perp
\ee
Now  if we set $ {\cal V}_\parallel= {\cal V}_\perp$, 
the excitation energies can also be made same,
 $ \bigtriangleup{\cal E}_\parallel= \bigtriangleup{\cal E}_\perp$, 
however the entanglement pressures do  always differ. 
We  calculate the entanglement asymmetry, in the  same way as \eqn{asym56},
\be\label{basym56}
{\cal A}_{nonconf} \equiv  
{\bigtriangleup S_\perp - \bigtriangleup S_\parallel
\over 
\bigtriangleup S_\perp +\bigtriangleup S_\parallel}
={\beta^2\gamma^2\over ( 2 + {\tp+3\over \tp-1}\beta^2\gamma^2)}\le {\tp-1 \over \tp+3}.
\ee
As discussed in the conformal case,
the bound gets saturated only in the case of D$p$-branes having wave like
 excitations at zero temperature. 
 For this we need to employ the same double limits 
$\beta\to 0,~~z_0\to\infty$, given in \eqn{dl2}, 
on the geometry  \eqn{bbst1}.  
Thus for nonconformal D-branes with a wave we  obtain the asymmetry ratio
\be\label{basym56a}
{\cal A}_{wave} \equiv  
 {\tp-1 \over \tp+3}.
\ee

In conclusion, our results assign 
maximum entanglement entropy asymmetry to the wave like excitations 
in a zero temperature CFT. The results can  be understood 
as we now eleborate. The wave like excitations 
in the CFT at zero temperature
generate finite entanglement pressure along the direction of
propagation of the wave, while the pressure remains
vanishing in all other (transverse) directions. 
When we switch on finite temperature in the CFT
(holographically including black hole in the bulk geometry) 
some entanglement pressure  gets distributed along the transverse
directions also. This finite temperature 
phenomenon reduces the net entanglement entropy asymmetry for the excitations. 
In the absence of a wave altogether the pressure becomes
 identical in all directions of the branes
 and hence entanglement asymmetry would also vanish.
Hence the asymmetry in entanglement entropy will necessarily exist 
if there are uniform
 wave like excitations or a uniform flow in the CFT. 
The asymmetry only gets amplified as temperature goes to vanishing values.

\section{Summary}

It has been shown that
the  entanglement pressure plays a significant role in determining the 
entanglement entropy for the strip  subsytems in the  CFT 
living on the boundary of  $AdS_{d+1}$ spacetime.
There is an entropy asymmetry along various directions of the  CFT
if their exists a pressure asymmetry. Besides the entropy asymmetry  
is directly proportinal to the pressure asymmetry. To quantify this
we  have determined  entanglement asymmetry ratio 
\be
{\cal A}\equiv  
{\bigtriangleup S_\perp - \bigtriangleup S_\parallel
\over 
\bigtriangleup S_\perp+\bigtriangleup S_\parallel}
={\beta^2\gamma^2\over ( 2 + {d+3\over d-1}\beta^2\gamma^2)}\le {d-1\over d+3}
\ee
which  depends only on the boost parameter 
$\beta$ and it is bounded. Interestingly
the bound is saturated in the large boost limit only \eqn{dl2}.
Thus a nonzero boost is simply a measure of the entanglement asymmetry. 
We have  discussed a large boost case which is the AdS-wave case.
Especially for the AdS waves 
there exist an optimum entanglement asymmetry 
\be
{\cal A}_{wave}={d-1\over d+3} 
\ee
which is a universal result at the first order in perturbation analysis. 
It is independent of any scale such as  
energy  of wave like excitations $\propto {1\over z_I^d}$.
We expect these results will  get corrected by higher orders of perturbation.

In the nonconformal D-branes cases the result gets slightly modified
\be
{\cal A}_{nonconf} 
={\beta^2\gamma^2\over ( 2 + {\tp+3\over \tp-1}\beta^2\gamma^2)}\le {\tp-1 \over \tp+3}.
\ee
The physical relevance of our results is indicated by the fact
that the entanglement entropy of subsytems is affected in the presence 
of boost, or a flow. It is not entirely an unexpected result 
as the boost indeed represents an asymmetric excitation of the system. 
It means subsystems along the flow and perpedicular to it get 
differently entangled as we have determined, 
$\bigtriangleup S_\perp > \bigtriangleup S_\parallel$. 
Upto first order this asymmetry is proportional to $\beta^2$ (for 
small velocities). These result  however will change at the second order 
 perturbative calculations.
Our results however imply more generic situations. 
Even in the absence of a flow, provided there exists  pressure 
asymmetry in the CFT due to some other reason, 
the entanglement asymmetry will always arise. 
The boosted black brane systems are used here only as the 
known examples to study
asymmetric systems. It would be worthwhile to explore other  systems 
like Bianchi models having more generic asymmetry.

\vskip.5cm
\noindent{\it Acknowledgments:} 
We are thankful to Arnab Kundu for the discussions.
\vskip.5cm
\appendix{
\section{The asymptotic expansion for nonconformal black D-branes}

The asymptotic expansion in the Fefferman-Graham  coordinates
is required to find the energy-momentum tensor of the boundary field theory.
The relevant details on holographic renormalization can be found in 
\cite{fg, kanit}. 
Let us define a new holographic coordinate $u$ through
\be
{z^2}=F^{-{4\over\tp}} u^2 , ~~~~~F=1+
{u^{\tilde p}\over u_0^{\tilde p}},~~~~
u_0^{\tilde p}\equiv 4 z_0^{\tilde p}
\ee
In these $u$ coordinates  an
expansion of \eqn{bbst1}
in the neighborhood of UV boundary $(u=0)$  becomes
\bea\label{bbst2}
&&ds^2\label{km1}
\simeq g_{eff}
\bigg[ {1\over u^2}\big[ (-1+ 4({\tp-1\over\tp}+\beta^2\gamma^2)
{u^{\tilde p}\over u_0^{\tilde p}}+\cdots) dt^2
+ (1+ 4({1\over\tp}+\beta^2\gamma^2)
{u^{\tilde p}\over u_0^{\tilde p}}+\cdots)dy^2 \br &&
~~~~~~~~~~
-{8\beta\gamma^2 u^4\over u_0^\tp} dt dy
+ (1+ {4\over\tp}
{u^{\tilde p}\over u_0^{\tilde p}}+\cdots) 
(dx_2^2+\cdots+dx_{p}^2)\big]+{4\over (5-p)^2}{du^2 \over u^2} +
d\Omega^2_{8-p}\bigg]\br &&
~~~~\equiv g_{eff}
\bigg[ {1\over u^2} (\eta_{\alpha\beta}+ t_{\alpha\beta}
u^{\tp}+ \cdots)dx^\alpha dx^\beta 
+{4\over (5-p)^2}{du^2 \over u^2} +
d\Omega^2_{8-p}\bigg]
\eea
The last line in the above equation indicates that 
 the spacetime geometry is expanded in  asymptotic neighborhood of
 conformally $AdS_{p+2}\times S^{8-p}$ spacetime. Besides in
these coordinates,
$u$ coincides with the energy scale of the $AdS_{p+2}$ geometry. The
$\eta_{\alpha\beta}$ is flat Minkowski
metric with index $\alpha=0,1,2,\cdots,p$. 
The  effective  coupling has the FG expansion (RG flow) given by 
\bea 
&& g_{eff}= {(\l_p u^{3-p})^{1\over 5-p}\over F^{3-p\over 7-p}}\simeq
(\l_p u^{3-p})^{1\over 5-p}
 (1- {3-p\over 7-p} {u^{\tilde p}\over u_0^{\tilde p}}+ \cdots)\br
\eea
In $p=3$ (conformal) case the $g_{eff}$ however remains fixed. 
The important point to notice from the FG expansion is
that the overall conformal factor  of the string
metric \eqn{km1}
and the string coupling $ e^\phi$ (given in \eqn{bbst1})  
are both  governed by the fluctuations of the single quantity $g_{eff}$. 
The fluctuations of the dilaton field, $\delta \phi$, can also be obtained 
from the expression
\bea 
&& e^{\phi}= {(2\pi)^{2-p}\over d_p N}
(\l_p u^{3-p})^{\tp\over 4}
 (1- {3-p\over 2} {u^{\tilde p}\over u_0^{\tilde p}}+ {\cal{O}}(u^{2\tp}))\br
&&~~~~\equiv e^{\phi_0}(1 + \delta \phi_{(\tp)} u^\tp + \cdots)
\eea
where $\phi_0$ represents the dilaton field in the absence of
the excitations. The first order fluctuation of dilaton are thus 
$\delta \phi_{(\tp)}=-{3-p\over 2} {1\over u_0^{\tilde p}}$. 
Obviously
$\delta \phi_{(\tp)}$ has opposite signs for $p>3$ and $p<3$ branes. (For
D$3$ brane $\delta \phi_{(\tp)}$  
vanishes as it should be for  $4D$ conformal field theory.)  
 The  nonvanishing  components of
stress-energy tensor of the
boundary  theory can now be obtained from the expression within the
angular brackets in  asymptotic expansion \eqn{bbst2}
\bea
&& t_{00}=({\tp-1\over \tp}+\beta^2\gamma^2) {4\over u_0^{\tilde p}}, ~~~
t_{yy}=({1\over \tp}+\beta^2\gamma^2) {4\over u_0^{\tilde p}} \br
&& t_{0y}=\beta\gamma^2 {4\over u_0^{\tilde p}}, ~~~
t_{ii}={1\over\tp} {4\over u_0^{\tilde p}},~~~(i=2,3,...,p)
\eea
The tensor $t_{\alpha\beta}$ has a nonvanishing trace. It is worthwhile to
observe   that the trace,
$t_{\alpha}^{~\alpha}$, and $\delta \phi$ have a  
relationship 
\be
{1\over4} t_\alpha^{~\alpha} -{3-p\over 7-p} \delta \phi_{(\tp)}=0
\ee
as they both  depend  
on  single deformation parameter  $u_0$. Actually this relation follows
from Ward identities in holographic renormalization of the
 boundary theory \cite{townsend}. Also 
$\nabla_\alpha t^{\alpha\beta}=0$ trivially.
We should not  be checking them over here as these are automatic 
in the FG expansion \eqn{km1} of nonextremal geometry.
The  energy 
of the excitations above the extremality for the boosted solutions is then
given by 
\bea 
\bigtriangleup {\cal E}= { V_p
\Theta_{8-p}Q_p\over 16\pi G_{N}}
 ({\tp-1\over\tp}+\beta^2\gamma^2)
{7-p\over z_0^\tp}
\eea
where $V_p$ is the $p$-dimensional spatial volume of all $x_i$'s and
$Q_p$ is a combinatoric factor defined earlier. 
 $\Theta_{8-p}$ is  unit  volume of the $S^{8-p}$,
 and $ G_{N}$ is the Newton's constant in ten dimensions. 
Similarly pressure components along the boost and in
 perpedicular directions are
\bea 
&&
\bigtriangleup{\cal{P}}_\parallel=
\bigtriangleup{\cal{P}}_y
= {
\Theta_{8-p}Q_p\over 16\pi G_{N}}
({1\over\tp}+\beta^2\gamma^2)
{7-p\over z_0^\tp}
\br &&
\bigtriangleup{\cal{P}}_\perp=
\bigtriangleup{\cal{P}}_{x_2}= {
\Theta_{8-p}Q_p\over 16\pi G_{N}}{7-p\over \tp z_0^\tp}=
\bigtriangleup{\cal{P}}_{x_3}= \cdots .
\eea
}

\vskip.5cm

\end{document}